# Electron Transport Through 2D Waveguide Using QTBM

Raja N. Mir and William R. Frensley

*Abstract—* Simulation of Electron Transport through two dimensional(2D) waveguide using Quantum Transport Boundary Method (QTBM) is done. Specifically, as an example the results of modeling L-shaped contact for a rectangular waveguide are presented. 2D-QTBM approach can be used in any scenario where the vertical axis of the device structure is significantly large and therefore the 2D approximation would be considered valid. This method elegantly generalizes to arbitrary shaped waveguides.

## I. Introduction

Electrons have both particle and wave nature. The wave nature of electron is taken into account by solving Schrodinger's equation. Several effects in the modern Metal Oxide Semiconductor(MOS) devices are explained taking into account the quantum mechanics. Examples include, electron tunneling through the thin oxide's, Gate Induced Drain Leakage (GIDL) or the quantization of the inversion layer in the Metal Oxide Semiconductor Field Effect Transistor (MOSFET) leading to an extra capacitance. Various properties of III-V based devices are also described well by taking into account the wave nature of electron. Many of these devices are opto-electronic in nature, e.g., Quantum Cascade Laser's (QCL's), infrared photodetectors, Light Emitting Diodes( LED's), tunnel FET's etc [4-10]. The simplifying assumption in case of these devices is usually that the devices are substantially thick so the plane vertical to the direction of transport is very large as compared to the device length and hence we can treat the transport as a one or two-dimensional problem. In case of the modern MOS transistors and at least 2D Schrodinger's equation must be solved to capture all the details relevant to the problem at hand. In the present work we apply these methods to solve the electron transport problem specifically for the L-Shaped contact in detail.

Raja N. Mir is Senior Member of IEEE and conducted the research in affiliation with University of Texas at Dallas. He currently works for NOKIA , 6000 Connection Dr, Irving TX 75039, USA (e-mail: raja.mir@nokia.com).
William R. Frensley is Fellow of IEEE and is with the School of Electrical Engineering, University of Texas at Dallas, 800 W Campbell Rd, Richardson, TX 75080, USA (e-mails: frensley@utdallas.edu).

## II. Mathematical Background

The single band QTBM [11-15] in 2D can be easily used to represent 2D quantum waveguides. The 2D systems are represented by linear combination of the localized states $|j, k>$, where j is the index of the position in the longitudinal direction and k is the index of orthogonal direction. The 2D multi-channel generalization can be done by casting the ψ, a and b as the sub-vectors and the matrix elements d, s, α and β as the sub-matrices indexed by k. Therefore ψ0, ψ1, ψn ... ψn+1 are all vectors in the transverse, k direction for a specific value of j. a and b are vectors representing the amplitudes of the modes of incoming and outgoing wavefunctions. The input and the output ports can be multiple in this formulation, in which case the constituent elements in the final matrix representation of the system become submatrices and vectors. The discrete form of Schrodinger's equation is [1-3] :

$$H\psi_j = -S_j\psi_{j-1} + D_j\psi_j - S_{j+1}\psi_{j+1} = E\psi_j \quad (1)$$

where H and E are Hamiltonian and Energy respectively, $D_j$ and $S_j$ are submatrices and $\psi_j$ is a vector. $\psi_j$ is next expanded on the basis of $\chi_{j;km}$ where m is the transverse mode. At the various boundaries we have, $\chi_{j+1;km} = z_{1;m}\chi_{j;km}$, z being the propagation factor for various transverse modes. Inserting this into (1) we have:

$$E\chi_{1;km} = \sum_{k'}\left[D_{1;kk'} - (z_{1;m} + z_{1;m}^{-1})S_{1;kk'}\right]\chi_{1;k'm} \quad (2)$$

Also,

$$\left|\psi \geq \sum_{jkm}(a_{1;m}z_{1;m}^{1-j})\chi_{1;km}\right|j,k> \quad (3)$$

At the points j = 0 and j = 1 we have:

$$\Psi_{0;k} = \sum_{m}(a_{1;m}z_{1;m}^{-1} + b_{1;m}z_{1;m})\chi_{1;km} \quad (4)$$

$$\Psi_{1;k} = \sum_{m}(a_{1;m} + b_{1;m})\chi_{1;km} \quad (5)$$

where a and b correspond to the amplitude of incoming wave and that of the outgoing wave. From these equations we can get the value of various a's in the following form:

$$a_1 = \alpha_1\psi_0 + \beta_1\psi_1 \quad (6)$$

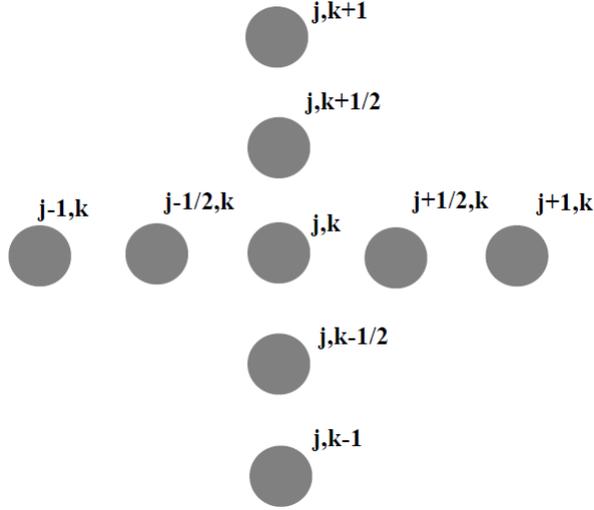

Fig. 1. The discretization scheme in 2D domain

$$a_n = \alpha_n \psi_{n+1} + \beta_n \psi_n \qquad (7)$$

Where α and β are matrices defined as:

$$\alpha_{1;mk} = \frac{1}{z_{1;m}^{-1} - z_{1;m}} [\chi_1^{-1}]_{mk} \qquad (8)$$

$$\beta_{1;mk} = -\frac{z_{1;m}}{z_{1;m}^{-1} - z_{1;m}} [\chi_1^{-1}]_{mk} \qquad (9)$$

$$\alpha_{n;mk} = \frac{1}{z_{n;m}^{-1} - z_{n;m}} [\chi_n^{-1}]_{mk} \qquad (10)$$

$$\beta_{n;mk} = -\frac{z_{n;m}}{z_{n;m}^{-1} - z_{n;m}} [\chi_n^{-1}]_{mk} \qquad (11)$$

In order to evaluate Schrodinger's equation numerically in the 2D case the 2D domain is divided into a grid of points, Fig.1. At each point the assumption is made that the wavefunction is related to only four points adjacent to it the top, bottom, right and left point.

Based on obtained roots of the Schrodinger's equation, we can have three categories: -
1. Real and positive: This type of root is associated with a decaying exponential form of wavefunction.
2. Real and negative: This type of root is associated with a decaying exponential with alternating signs form of wavefunction.
3. Complex conjugate: This type of root is associated with a propagating wavefunctions.

### III. RESULTS AND DISCUSSION

We have simulated a 2D structure using 2D-QTBM. The structure modelled here has rectangular shape. With 30 atomic layers in one axis and 10 atomic layers in the orthogonal axis. Electrons are launched from one end and get out through the other end in the longitudinal direction. In the transverse directions there is no transport. When we introduce L shaped contacts and we stop the transmission of the electron in the longitudinal direction where the contact is present. We have assumed the contacts to mean non-transmitting Schottky boundaries. The structure is divided into 4 boundaries. II lead launches the wave into the structure using the QTBM boundary condition. IV is the output lead. I and III are the boundaries which have no transport through them and outside these boundaries the wavefunction goes to zero and these give rise to sinusoidal modes at the input of the structure in the transverse direction. We can choose to launch the wave in any one of these modes.

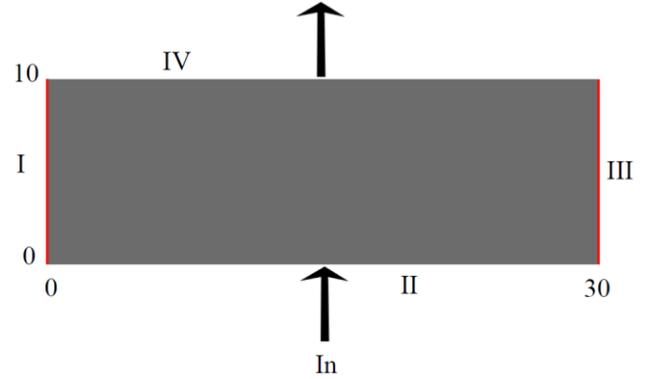

Fig. 2. Structure simulated is rectangular with two transmitting boundaries and two boundaries which are closed. The wave comes in from one face and leaves from another.

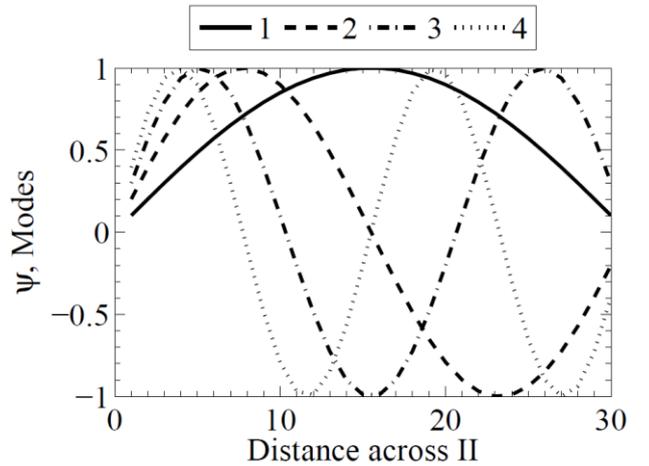

Fig. 3. The sections II and IV in the previous figure are closed on two sides and hence the wavefunction must go to zero right outside these two nodes. Hence these two ends will support sine like wave functions in these directions(transverse). Which will be used as the basis functions.

higher order mode in transverse direction the share of energy is more and the longitudinal mode becomes evanescent.

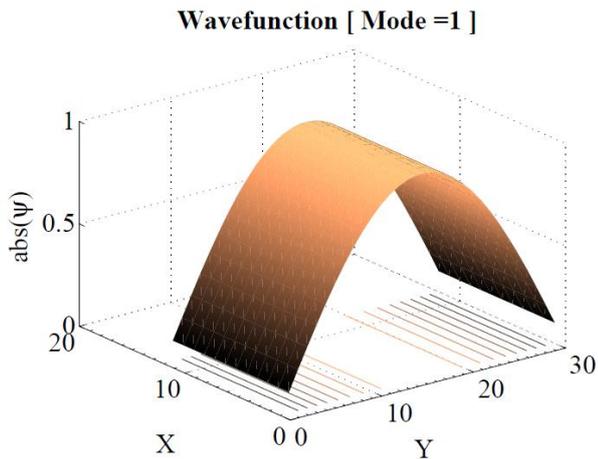

Fig. 4. Here a wave is launched with E=0.1 eVs and showing the first mode going through the structure. The wave is going in from the side the reader is looking at the figure or in other words we are looking at side II of the structure. The length of contact on the IV end is half of the length of the IV length i.e: 15 nm.

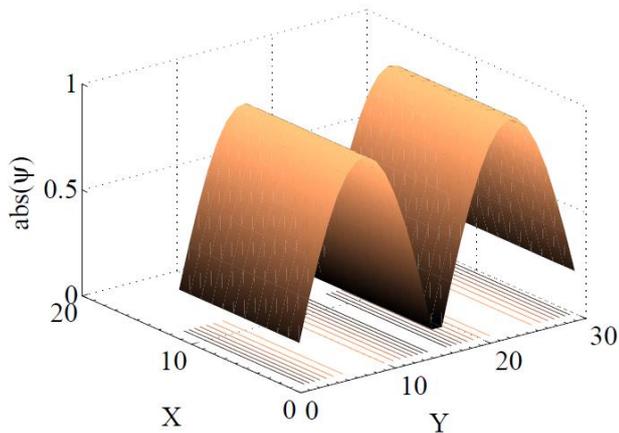

Fig. 5. A wave launched with E=0.1 eV's and showing the second mode going through the structure.

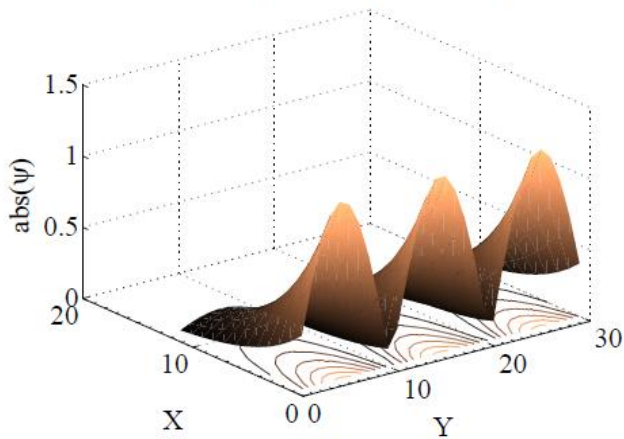

Fig. 6. A wave launched with E=0.1 eVs and showing the third mode going through the structure. We can observe that the third mode is evanescent which is because of the fact that the energy that we launch the wave with is shared between the transverse and the longitudinal modes and since we have excited

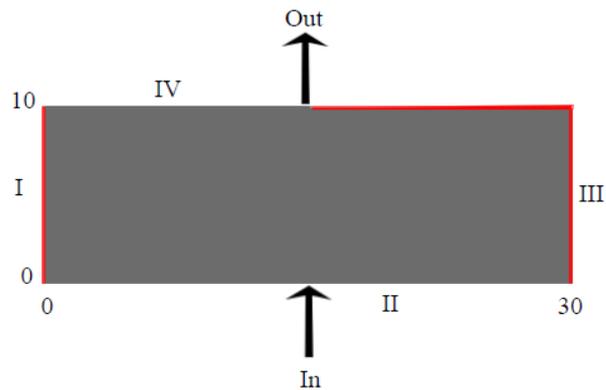

Fig. 7 An L-shaped contact is represented. We represent contacts by red lines. The length of the contact on face IV is adjustable such that we can see the effects of the transmission of the electrons through the structure for various contact lengths. The open boundary conditions are formulated for the non-colored regions. Notice our contacts are non-transmitting Schottky contacts.

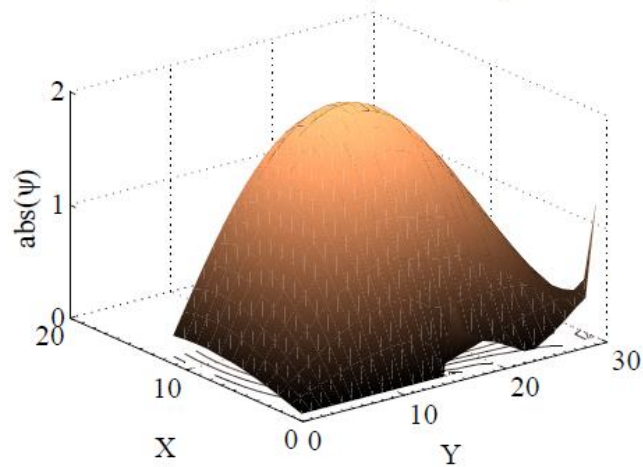

Fig. 8 Here a wave is launched with E=0.1 eVs and showing the first mode going through the structure. The wave is coming out from the side the reader is looking at the figure or in other words we are looking at side IV of the structure. The length of contact on the IV end is half of the length of the IV length i.e: 15 nm.

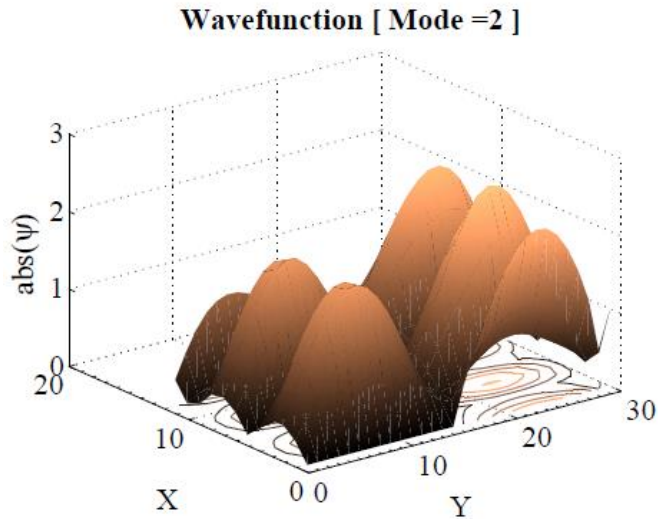

Fig. 9 Here a wave is launched with E=0.1 eVs and showing the second mode going through the structure. The wave is coming out from the side the reader is looking at the figure or in other words we are looking at side IV of the structure. The length of contact on the IV end is half of the length of the IV length i.e: 15 nm.

## IV. CONCLUSION

Many effects present in the modern nanoscale electron devices can be captured if the quantum mechanics is formulated correctly. In this paper we have presented the results of modelling a 2D electron waveguide using 2D Quantum Transmitting Boundary Method. It is shown that the 2D Quantum Transmitting Boundary Method is effective in simulating arbitrary boundary conditions very efficiently. These simulations can be applied to cases where the transverse direction is large and uniform so that the 2D approximation of the semiconductor structure holds.